\documentclass[a4paper]{article}
\usepackage{speechprosody2012,amssymb,amsmath,epsfig}
\usepackage[french]{babel}
\usepackage[utf8]{inputenc}

\usepackage{graphicx}
\usepackage{array}
\usepackage{colortbl}
\usepackage{rotating}
\usepackage{makeidx}
\usepackage{multirow}
\usepackage{textcomp}
\usepackage{psfrag}
\usepackage{makeidx}
\usepackage{hyperref}
\usepackage{enumitem}
\usepackage{booktabs}
\usepackage{subcaption}
\graphicspath{ {figures/} }
\usepackage{array}
\usepackage[list=true,listformat=simple]{subcaption}

\usepackage{hyperref}
\hypersetup{urlbordercolor={0 1 1}}

 \def\reg{{\rm\ooalign{\hfil
      \raise.07ex\hbox{\scriptsize R}\hfil\crcr\mathhexbox20D}}}

\usepackage{hyperref}
\graphicspath {{./figures/}}

\newcounter{mainmemorder}

\title{Att-HACK: An Expressive Speech Database with Social Attitudes}

\makeatletter
\def\name#1{\gdef\@name{#1\\}}
\makeatother
\name{{\em Clément Le Moine, Nicolas Obin }}
\address{ STMS Lab - IRCAM, CNRS, Sorbonne Université \\ Paris, France}

\begin{document}

\ninept
\maketitle
\shorthandoff{:}

\vspace{-0.5cm}

\begin{abstract}

This paper presents Att-HACK, the first large database of acted speech with social attitudes. Available databases of expressive speech are rare and very often restricted to the primary emotions: anger, joy, sadness, fear. This greatly limits the scope of the research on expressive speech. Besides, a fundamental aspect of speech prosody is always ignored and missing from such databases: its variety, i.e. the possibility to repeat an utterance while varying its prosody. This paper represents a first attempt to widen the scope of expressivity in speech, by providing a database of acted speech with social attitudes: friendly, seductive, dominant, and distant. The proposed database comprises 25 speakers interpreting 100 utterances in 4 social attitudes, with 3-5 repetitions each per attitude for a total of around 30 hours of speech. The Att-HACK is freely available for academic research under a Creative Commons Licence.

\end{abstract}

\vspace{0.25cm}

\renewcommand{\thefootnote}{\fnsymbol{footnote}}


\section{Introduction}
\label{sec:intro}

Though the linguistic functions of speech prosody are nowadays well documented in a large number of languages (phonology, syntax/prosody interface, etc...), its expressive or para-linguistic functions, such as speaking style or speech emotions, does not benefit from the same amount of attention from the linguistic community.
Meanwhile, speech engineers have realized spectacular advances in the past decade creating extremely realistic synthetic voices \cite{wang17_tacotron} which are now integrated into voice interfaces that are increasingly present in our everyday lives, such as the voice assistants and conversational/virtual agents. However, these intelligible and natural voices still clearly lack expressiveness and adaptability which greatly limits the interaction between humans and machines (see for instance \cite{Cas12}). 
Expressivity is the next frontier of speech research at the interface of linguistics and technology, as shown by the recent increase of research in this domain \cite{wang2018style}. Consequently, there is a clear need to better understand the expressivity of the human voice which is mainly conveyed through speech prosody. In particular, there is a growing interest of the engineering community in the statistical modelling of prosody \cite{Lat08, Obi11c, Yin16, Wan17, Luo2017-qn, Obi18, Gerazov2018b}, expressive speech synthesis \cite{ObinPhD, Eyben12, Akuzawa19}, and conversion of neutral to expressive speech \cite{Veaux2011-pq, Rob01, Luo_2019_0}. \\

Nowadays, speech expressivity is generally equated to speech emotions though the scope of expressivity includes but is not restricted to the primary emotions as denoted by Ekman \cite{Ekman92}. This limitation is probably due to the difficulty of converging to an agreement on the terminology used to describe the various and subtle forms of expressivity in speech.
Accordingly, the study of speech expressivity is generally limited to dedicated speech emotion databases as interpreted by actors or to audio books read by professional readers (mostly in English, and sometimes in French or German \cite{Chen}). In the past decade, speech emotion research has mainly focused on acted emotional speech: from its original form in which an actor is asked to interpret a short text with a given emotion \cite{Bur05} to more open and spontaneous forms in which two actors freely improvise based on a given scenario and then asked to rate their own speech emotions with categories or on valence/arousal continuous scales \cite{Bus08, Keo10}.  Moreover, these databases have been created only for the purpose of speech emotion recognition, but not for the modelling of speech prosody and neither for speech synthesis and conversion. \\

However, speech expressivity is not restricted to primary emotions. For instance, to deal with more subtle variation, the circumplex model, in which emotions are categorized in a bidimensional space, was proposed by psychologists \cite{russel80}. Attitude was firstly equated to the first dimension (valence) of this model \cite{ajzen80}. A distinction between emotion and attitude was done in \cite{Couper} by defining emotion as a speaker state and attitude as a kind of behaviour. This distinction was later refined in \cite{Wic00}, by defining attitude as a predictor of social behaviour. The attitudinal aspect of expressiveness of course differs from the primary emotions, and create a distinction between the propositional attitude (towards an utterance: irony, doubt, etc... \cite{Lee83}) and the social attitudes (towards a person: dominant, friendly, seductive, distant, etc...). This last dimension has been recently investigated in the study of the role of speech prosody in neurosciences \cite{Pon18}. Moreover, all of the existing speech databases miss an essential aspect of speech prosody: its {\em variety} \cite{Obi12}. There is always only one realization of each utterance, while any utterance can be obviously realized with many prosodies, some being functionally equivalent, some having various degrees of expressivity. There is a clear need for speech database that would tackle these issues and allow a diversification of the research in speech prosody and expressivity, and with sufficient data to allow learning generative models. \\

This paper presents Att-HACK, the first large database of acted speech with social attitudes. This database comprises the recordings of 25 speakers (M/F) interpreting 100 utterances in 4 different social attitudes: friendly, seductive, dominant, distant.
In a given attitude, each utterance is repeated differently between 3 and 5 times in order to provide access to the inherent prosodic variety of speech.
This represents a total of about 30 hours of speech, which comes with audio signals, orthographic text transcription, F0 analysis, and phonetic alignments. The Att-HACK \footnote{\url{http://www.openslr.org/88/}} is freely available for research under a Creative Commons (BY/NC/ND) Licence. The remaining of the paper is organized as follows: in Section \ref{sec:conception} we describe the conception of Att-HACK. A full description of the database is proposed in Section \ref{sec:analysis}, with some illustrations of the F0 obtained in the 4 social attitudes by some of the speakers.

\section{Conception of Att-HACK}
\label{sec:conception}

In this Section, the full conception of the Att-HACK speech database of social attitudes is described, from the initial choice of social attitudes as a continuation of speech expressivity research and including the details of the speech database design.

\subsection{Models of Emotions}

The means by which humans construct representations of emotions is still an open and debated issue in the field of psychology and affective sciences: today, two fundamental models are used for the description of emotions. The first approach is \emph{categorial} and supports the idea that emotions are discrete and relate to concepts that are essentially distinct. A discrete emotion theory was developed by Ekman et al. in 1992 \cite{Ekman92} in which six basic emotions were proposed: anger, disgust, fear, happiness, sadness, and surprise. The second approach is dimensional and supports that emotions are defined according to at least one dimension. In the circumplex model \cite{russel80}, emotion is defined on two dimensions: valence (positive–negative) and arousal (passive-active). A third dimension denoted dominance (submissiveness-dominance) was added to compose the PAD (Pleasure, Arousal, Dominance) three-factors model for speech emotions \cite{Russ01}.

\subsection{Definition and Choice of Social Attitudes}

The choice of social attitudes is methodologically rooted with the idea of extending the study of expressivity from primary emotions to more subtle forms. As initially inspired by \cite{Wic00}, social attitudes allows to represent the attitude of a speaker towards its interlocutor during an interaction. Such social attitudes differ from emotions which are internal state of a speaker and from propositionnal attitudes which are the attitude of a speaker towards an utterance. 

\begin{figure}[h!]
    \centering
    \includegraphics[scale=0.49]{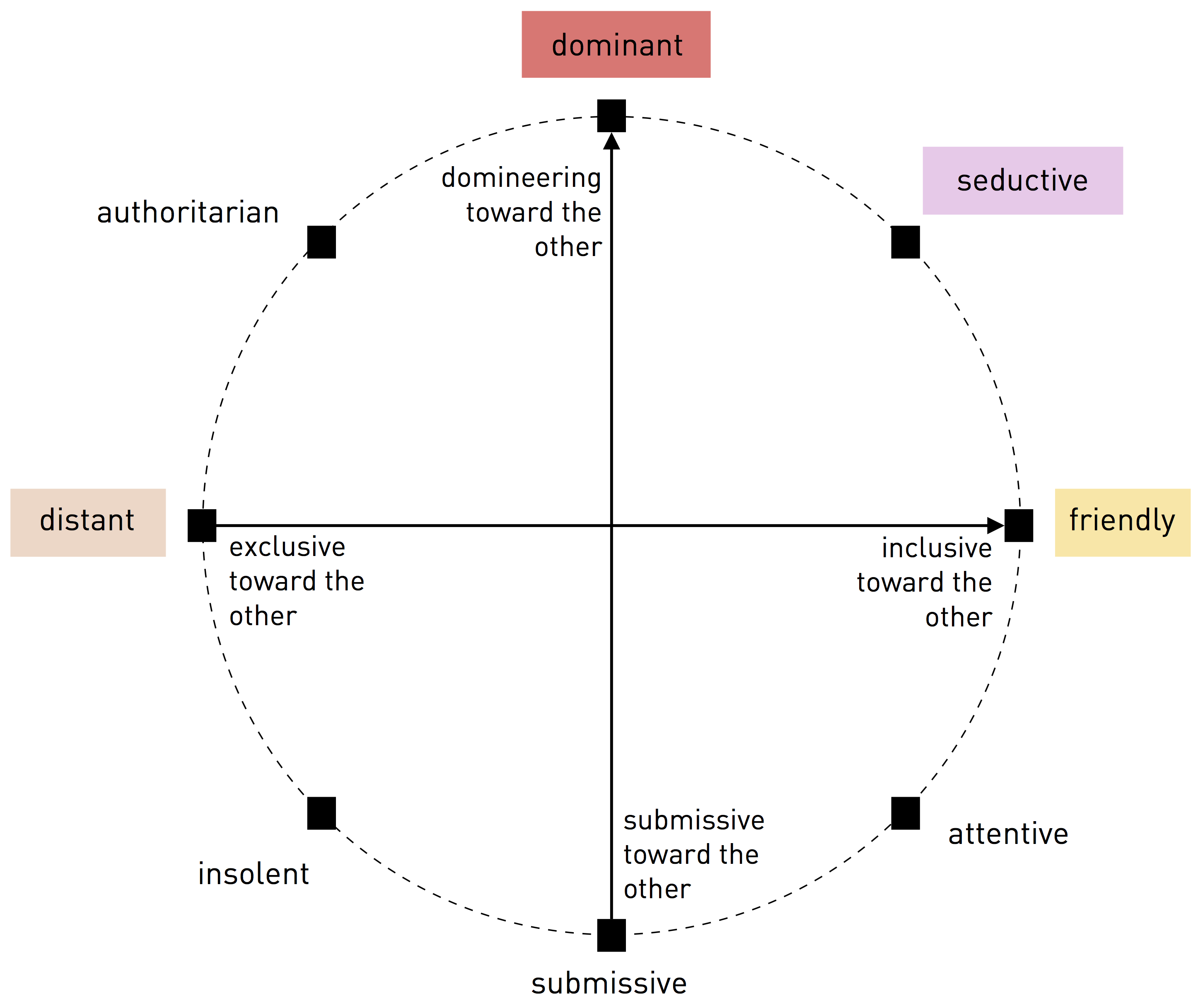}
    \caption{Social attitudes represented in a 2-D space inspired by Jean Julien Aucouturier's categorization for musicians playing attitudes \cite{auc17}.}
    \label{fig:att}
\end{figure}


\vspace{1cm}

Following this initial definition, Jean-Julien Aucouturier et al. \cite{auc17} recently proposed an original categorization for musicians' playing attitudes inspired by Leary's rose model \cite{Leary57}. The attitudes are described in a bidimensional space, the first dimension reflects the hostility/friendliness (alternatively, negative/positive) towards the other musician while the second dimension reflect the position of the musician in a social hierarchy (subordination/dominance). 
Our proposed choice of social attitudes is in the continuation of these two precursory works. In this paper, four social attitudes were defined by sampling the valence and dominance dimensions during a speech interaction: friendly, seductive, dominant and distant (figure \ref{fig:att}). This includes one negative (distant) and three positives (friendly, seductive, dominant) attitudes sampled in the semi-space of neutral to high-hierarchy in the dominance space. \\

\noindent The four attitudes are described as follows :

\begin{itemize}
    \item \textbf{friendly}: you are pleasant and benevolent, you care about others' preferences, you act towards the others independently from your own situation.
    \item \textbf{seductive}: everything in your behaviour aims at charming the others, to make them love you, you do not care about others' preferences but you are ready for anything to seduce them even if you have to fake benevolence.
    \item \textbf{dominant}: you are self confident, sure of your own superiority, you do not care about others' preferences, everything in your behaviour is dedicated to make the others obey and listen to you without imposing anything explicitly.
    \item \textbf{distant}: you (barely) do not care about the others, you are uncommunicative, you do not care about others' preferences.
\end{itemize}

\begin{figure}[h!]
    \centering
    \includegraphics[scale=0.5]{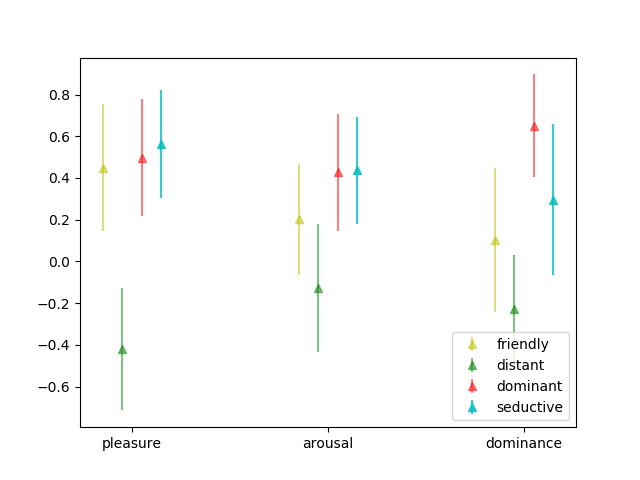}
    \caption{Social attitudes represented with the PAD model proposed by Russel \& Mehrabian \cite{Russ01}}
    \label{fig:dom}
\end{figure}

An attempt to describe our four attitudes through using the Russell's PAD model has been initiated by recomposing those attitudes with some of the 151 emotion-denoting terms subjects were asked to rate in \cite{Russ01} using valence, arousal, and dominance. Let us assume that our \text{friendly} is a mix of $\{$friendly, humble, curious, respectful, kind$\}$, \textit{seductive} a mix of $\{$affectionate, sexually excited, controlling$\}$, \textit{dominant} a mix of $\{$influential, domineering, bold$\}$ and \textit{distant} a mix of $\{$timid, bored, inhibited, uninterested, detached, shy, snobbish lonely$\}$. The figure \ref{fig:dom} shows our four social attitudes categorized according to the three Russel's dimensions (by averaging of the results obtained by Russel for the considered groups).

\subsection{Set of sentences}

The set of sentences used for the creation of the database has been designed in French as inspired by the corpus of propositionnal attitudes proposed by Morlec in \cite{Mor97}. The proposed sentences have been designed according to the following criteria: 1) to avoid introducing a bias due to the expressive content of the text, the sentences were designed to be as neutral as possible; 2) to reduce the prosodic variability due to the text structure, the sentences were designed with a limited and controlled linguistic complexity, by using simple syntactic structures and short sentences; 3) Finally, the  sentences were designed in a way they remain plausible in each social attitude.  

\begin{figure}[h!]
    \centering
    \includegraphics[scale=0.5]{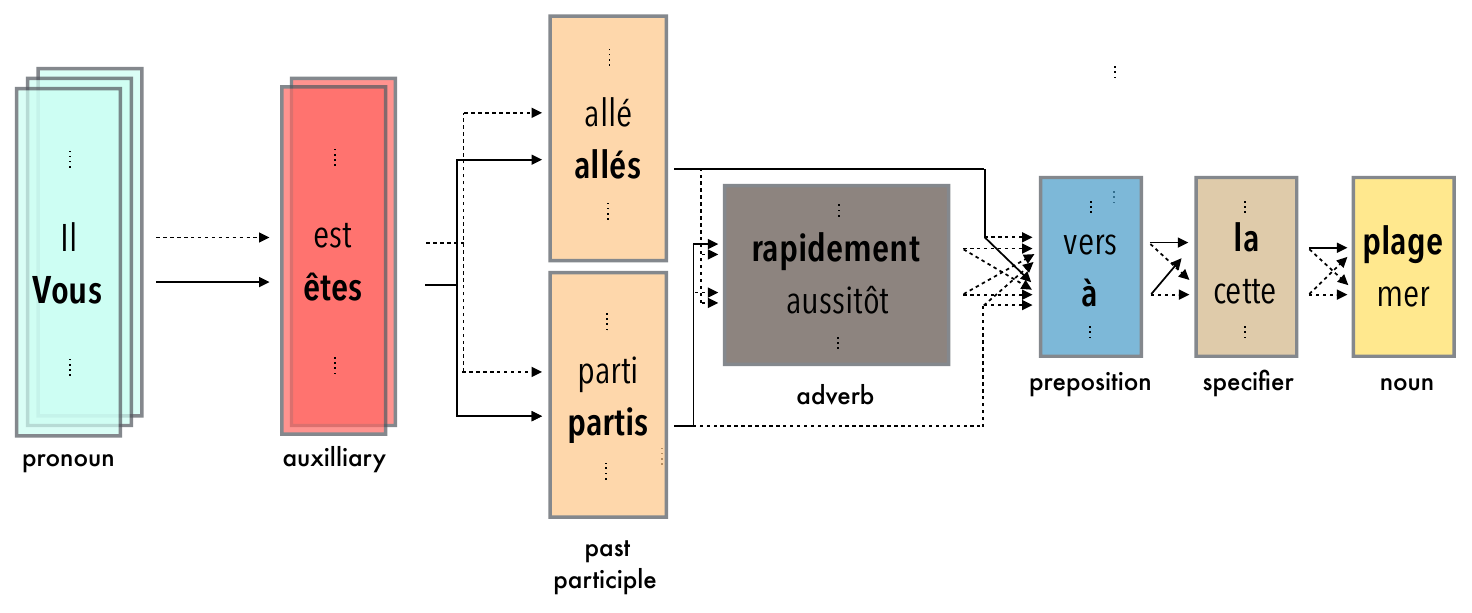}
    \caption{Phrase generator functioning for the above quoted phrases, chosen words are in bold, dotted lines represent possible choices for the algorithm}
    \label{fig:gen}
\end{figure}

Accordingly, we constructed a set of 100 sentences (from 2 to 8 syllables) corresponding to simple everyday life situations  (classic situations in classic socialization places like home, restaurant, workplace, ...).  For this purpose, we designed a phrase generator that builds phrases from semantic nucleus (\{pronoun/noun + verb\} or \{pronoun/noun + auxilliary\}) by randomly picking words in dictionaries in order to guarantee the phrases are always conceived the same way. We randomly kept 100 phrases among the 10000 generated ones to build our set of sentences, a sample of 10 sentences is listed below. 

\begin{center}
    \emph{Oui} \\
    \emph{Bonjour} \\
    \emph{C’est vrai} \\
    \emph{A demain Paul} \\
    \emph{Bonne journée Marie} \\
    \emph{Il est tard à Londres} \\
    \emph{Vous êtes allés à la plage} \\
    \emph{Vous êtes partis rapidement} \\
    \emph{Impossible, attendons un peu} \\
    \emph{C’est vrai, allons prendre un café} \\
\end{center}

Figure \ref{fig:gen} depicts the functioning of the automatic sentence generator considering the example of two sentences created from the nucleus ‘‘Vous êtes ..." (‘‘You are ...'').

\section{Description of Att-HACK}
\label{sec:analysis}

\subsection{Audio recordings}

To feed this database, twenty-five actors were recorded in professional studios at Ircam. It consisted of 4 hours sessions during which one actor had to play 100 sentences in the four different social attitudes, proposing from 3 to 6 different versions of each sentence in each attitude. At the beginning of each session, the four attitudes were shortly described as stipulated above. Those descriptions have been used as acting options, actors were told to act the way they felt regarding each attitude denomination and not necessary in regards to those descriptions. The actors were told to be as natural as possible, no other information was given during the session. \\

For the recording, we used a Neumann U87 static microphone plugged into a RME fireface interface synchronized with the ProTools software. All audio recordings were made with a sampling rate frequency of 44.1 kHz and quantization of 16 bits per sample. The recording were made in two different studios depending on their availability. A patch implemented with the Max for Live software \footnote{\url{https://www.ableton.com/en/live/max-for-live/}} was used to provide a visual interface to the actor, displaying on a screen in front of the actor the sentence to be read and the expected attitude, this display being monitored by a sound engineer. The patch also allowed to store the time codes corresponding to each sentence, which were used after the session to segment and name automatically the continuous recording made with the ProTools software. At the end of a session, we had at least 2500 audio files for each actor. This amount of recordings has been manually sorted to remove corrupted files and utterances that were judged as being not natural enough or badly acted. At the time of writing, the Att-HACK database is composed of more than 22,000 expressive utterances, which will be completed in the months to come to reach around 50,000 utterances and 30 hours of speech. 

\subsection{Speech to text alignment}

Speech-to-text alignment was performed by using the ircamAlign software \cite{Lan08} which is based on the HTK toolbox and the Lia\_phon French phonetizer and learned on the BREF French database of read speech. Phonetic alignment is first processed by ircamAlign and followed by rule-based syllable segmentation and a simple phrase segmentation. The resulting alignment are stored in dedicated lab files, a text format indicating at each line the starting time, ending time, and label corresponding to the text sentence and audio recording.  

\begin{figure}[h!]
    \centering
    \includegraphics[width=0.39\textwidth]{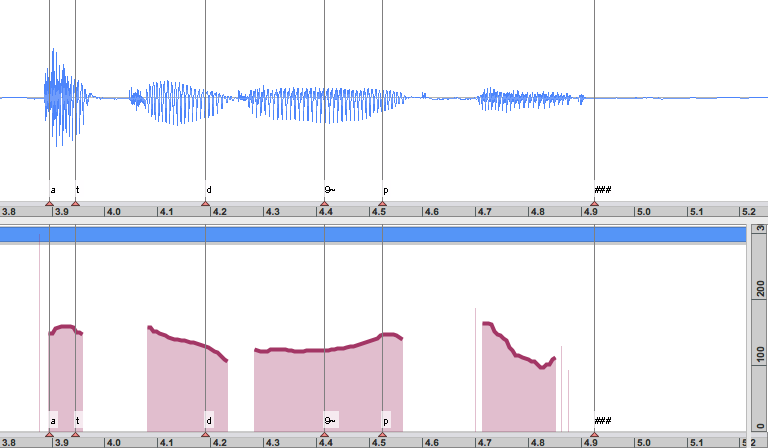}
    \caption{Syllable segmentation (above) and F0 values as estimated on voiced frames (below)}
    \label{fig:sig_f0}
\end{figure}

\newpage

\subsection{F0 estimation}

The fundamental frequency of the speakers was estimated by using the SWIPEP algorithm \cite{Cam07} with a minimum pitch value of  $75$ Hz, a maximum pitch value of $450$ Hz, and a hop size of $5$ ms, without any post-processing for correcting or smoothing the raw pitch values. The voiced/unvoiced decision was computed from the pitch strength associated with the pitch value estimate with a threshold of $0.25$ (the pitch strength being a value between 0 and 1 corresponding to the periodicity of the speech frame).

Figure \ref{fig:sig_f0} presents the phonetic segmentation of the speech waveform (above) in syllables and the corresponding F0 values (below) for the sentence \textit{"Attendons un peu"} using the Audiosculpt software \footnote{\url{https://forum.ircam.fr/projects/detail/audiosculpt/}}.

\section{Preliminary investigation}
\label{sec:investigation}

A preliminary investigation was conducted to illustrate the content of the Att-HACK speech database of social attitudes.

\subsection{Extraction of F0 contours}

A F0 contour was extracted for each syllable in order to illustrate the F0 patterns realized by the actors for the different social attitudes. The F0 contour of a syllable was identified as the one corresponding the longest sequence of F0 values considered as voiced over the syllable, i.e. the longest F0 segment for which each F0 value corresponds to a pitch strength value which is above a given  threshold (in this paper, $0.3$).

\begin{figure}[!h]
    \centering
    \begin{subfigure}[b]{0.48\textwidth}
        \includegraphics[width=\textwidth]{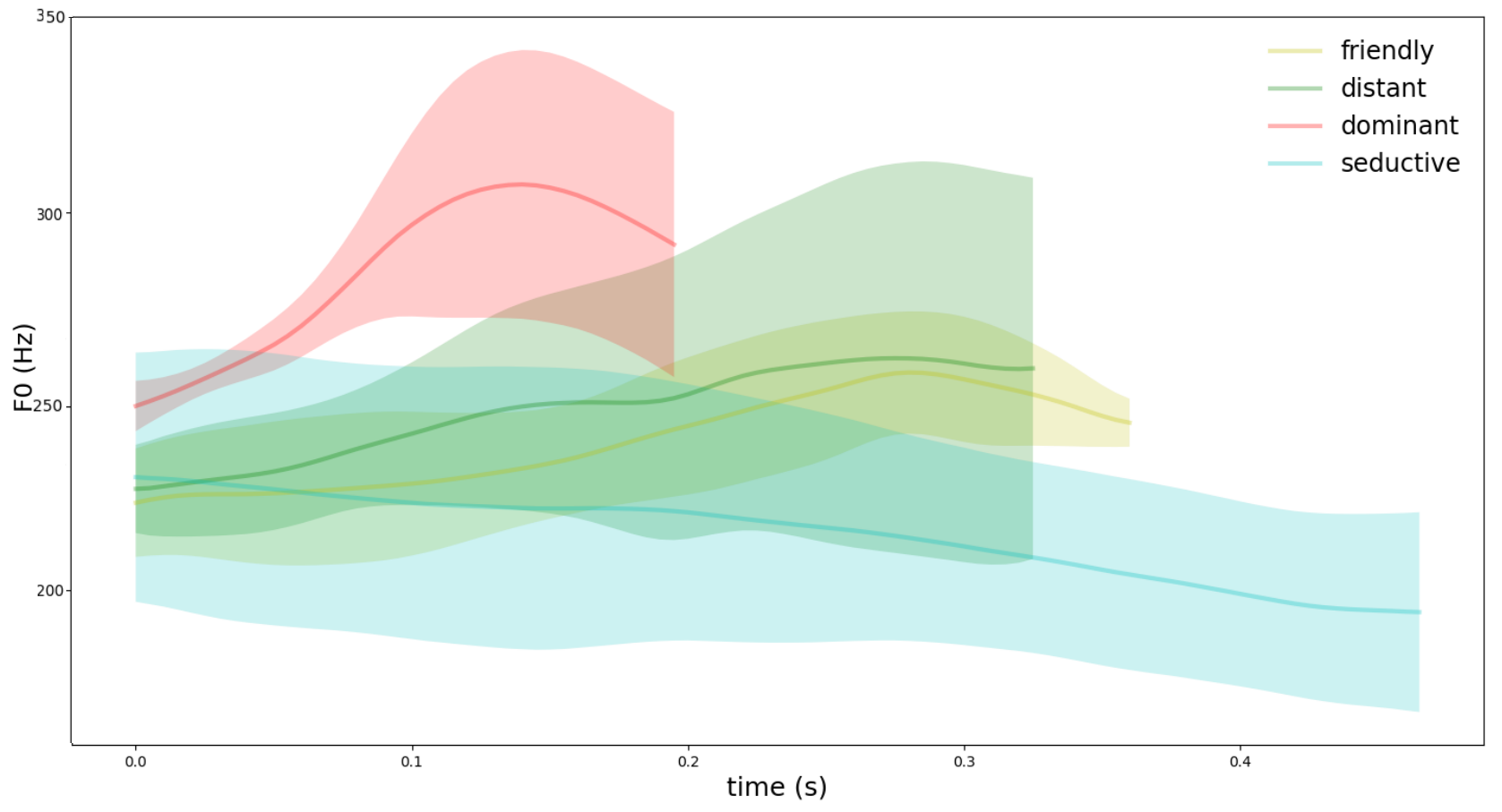}
    \end{subfigure}
    \begin{subfigure}[b]{0.48\textwidth}
        \includegraphics[width=\textwidth]{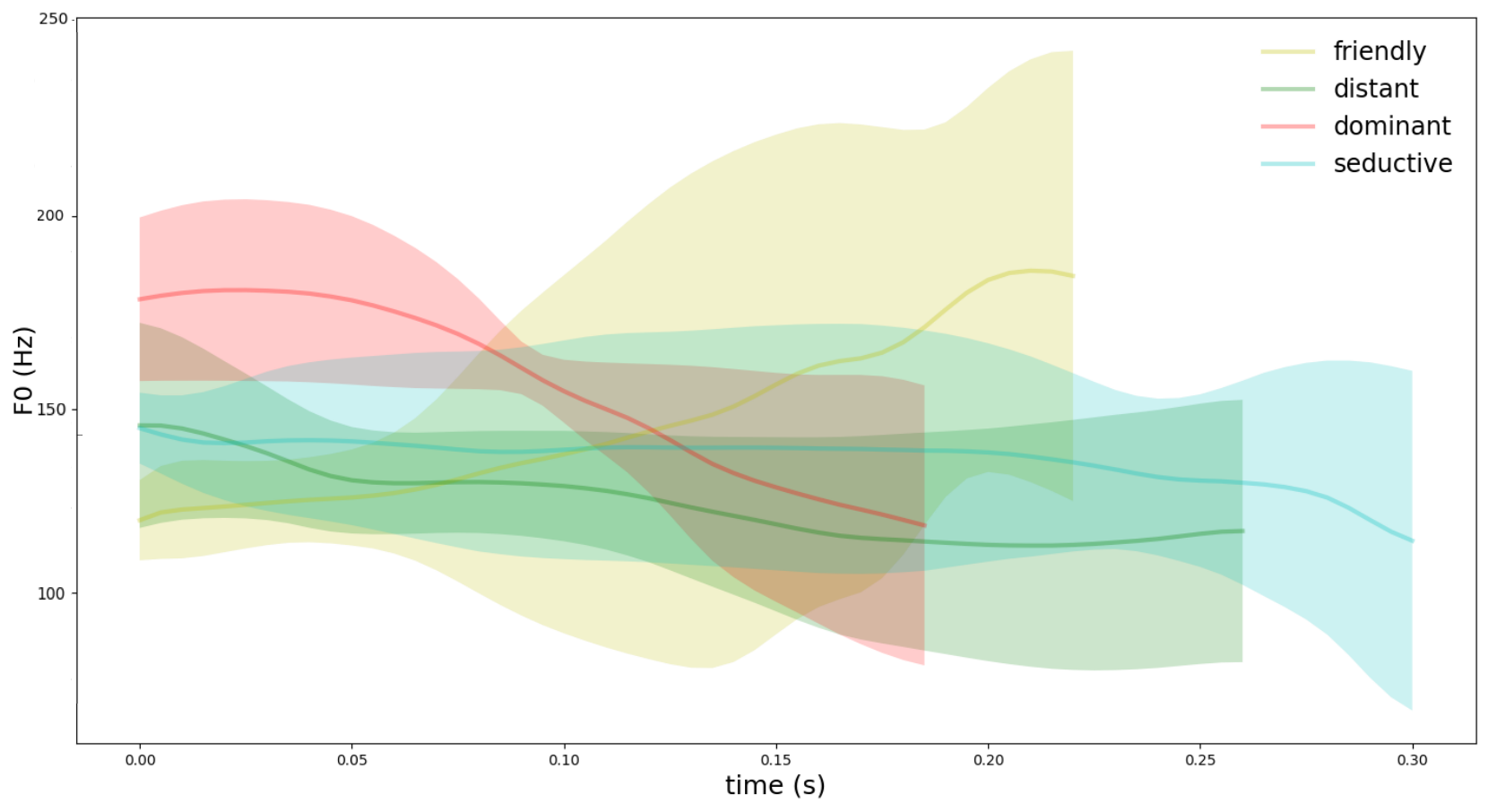}
    \end{subfigure}
    \caption{F0 contours mean (solid line) and standard deviation (filled with color) for the phrase \textit{"Oui"} for a female speaker (above) and a male speaker (below)}
    \label{fig:contours}
\end{figure}

\subsection{F0 Statistics vs Attitudes}
 
 A preliminary investigation was conducted to compare the F0 statistics of the actors across the social attitudes.  We computed mean and standard deviation statistics on F0 segments extracted as stipulated above, for each speaker and attitude. These statistics are reported in Table \ref{tab:means_and_stds}, including global statistics (female, male and mixed-gender). It is to be noted that only the first part of the recordings has yet been post processed at the time of writing. \\
 
Figure \ref{fig:contours} illustrates F0 pattern distributions obtained for a given sentence with the four attitudes, each attitude being represented by a dedicated color. In each color, the solid line represents the F0 pattern obtained by averaging the variations realized by the actor, the area filled with color represents the corresponding standard deviation, and the length of the pattern the corresponding mean duration. This illustration reveals that distinctive F0 patterns are associated with the social attitudes, and also highlights the diversity of strategies employed by actors to communicate a social attitude. 

\begin{table}[tb]\centering
\scalebox {0.75}[0.75]{
\begin{tabular}{p{1cm}p{1cm}cccccccc}
\hline
\textbf{gender} & \multicolumn{2}{c|}{\textbf{Friendly}} & \multicolumn{2}{|c|}{\textbf{Seductive}} & \multicolumn{2}{|c|}{\textbf{Dominant}} & \multicolumn{2}{|c}{\textbf{Distant}} \\ 

 \cline{2-9}
    & Mean & Std & Mean & Std & Mean & Std & Mean & Std\\
\hline
\textit{female} & 208 & 11 & 186 & 12 & 207 & 10 & 186 & 10\\
\textit{male} & 112 & 8 & 113 & 10 & 119 & 8 & 104 & 7\\
\hline
\textit{global} & 160 & 10 & 150 & 11 & 163 & 9 & 145 & 9\\
\hline
\end{tabular}}
\caption{Syllable F0 contours means and standard deviations (in Hz) for female and male speakers of Att-HACK}
\label{tab:means_and_stds}
\end{table}

\begin{table}[tb]\centering
\scalebox {0.75}[0.75]{
\begin{tabular}{p{1cm}p{1cm}cccccccc}
\hline
\textbf{gender} & \multicolumn{2}{c|}{\textbf{Friendly}} & \multicolumn{2}{|c|}{\textbf{Seductive}} & \multicolumn{2}{|c|}{\textbf{Dominant}} & \multicolumn{2}{|c}{\textbf{Distant}} \\ 

 \cline{2-9}
    & Mean & Std & Mean & Std & Mean & Std & Mean & Std\\
\hline
\textit{female} & 404 & 151 & 426 & 141 & 399 & 143 & 440 & 169\\
\textit{male} & 410 & 133 & 440 & 145 & 413 & 137 & 471 & 177\\
\hline
\textit{global} & 407 & 144 & 431 & 142 & 405 & 141 & 452 & 172\\
\hline
\end{tabular}}
\caption{Syllable durations means and standard deviations (in ms) for female and male speakers of Att-HACK}
\label{tab:means_and_stds}
\end{table}

\section{Conclusion}

This paper presents Att-HACK, a first attempt to widen the scope of expressivity in speech, by providing a database of acted speech with social attitudes: friendly, seductive, dominant, and distant. The proposed database comprises 25 speakers interpreting 100 utterances in 4 social attitudes, with 3-5 repetitions each per attitude proving a great prosodic variety for a total of around 28 hours of expressive speech. The Att-HACK is freely available for academic research under Creative Commons Licence.

\section{Acknowledgements}

This research has been supported by the French Ph2D/IDF MoVE project on MOdelling of speech attitudes and application to an expressive conversationnal agent and funded by the Ile-De-France region and by the French ANR project TheVoice: ANR-17-CE23-0025.


\bibliographystyle{IEEEtran}
\bibliography{ref}
\end{document}